\documentclass[conference]{IEEEtran}
\IEEEoverridecommandlockouts

\usepackage{cite}
\usepackage{amsmath,amssymb,amsfonts}
\usepackage{algorithm}
\usepackage{algorithmicx}
\usepackage{algpseudocode}
\algdef{SE}[SUBALG]{Indent}{EndIndent}{}{\algorithmicend\ }%
\algtext*{Indent}
\algtext*{EndIndent}
\usepackage{lipsum}

\usepackage{threeparttable} 
\usepackage[export]{adjustbox}
\usepackage[table]{xcolor}
\usepackage{url}
\usepackage{bm}
\usepackage{courier}
\usepackage{bbm}
\usepackage{caption}
\usepackage{subfig}

\usepackage{array}
\usepackage{multirow}
\usepackage{graphicx}
\usepackage{tablefootnote}
\usepackage[T1]{fontenc}
\usepackage{booktabs}
\usepackage{multirow}
\usepackage{hyperref}

\DeclareMathOperator*{\argmin}{arg\,min}

\usepackage{textcomp}
\usepackage{xcolor}
\def\BibTeX{{\rm B\kern-.05em{\sc i\kern-.025em b}\kern-.08em
    T\kern-.1667em\lower.7ex\hbox{E}\kern-.125emX}}
\begin{document}

\title{Differentiable Quantum Architecture Search for Quantum Reinforcement Learning\\
}

\author{\IEEEauthorblockN{Yize Sun\IEEEauthorrefmark{1},
Yunpu Ma\IEEEauthorrefmark{2}, Volker Tresp\IEEEauthorrefmark{3}}
\IEEEauthorblockA{Ludwig Maximilians University\\Siemens AG\\
Munich, Germany\\
Email: \IEEEauthorrefmark{1}yize.sun@campus.lmu.de,
\IEEEauthorrefmark{2}
cognitive.yunpu@gmail.com,
\IEEEauthorrefmark{3}volker.tresp@lmu.de}}


\maketitle

\begin{abstract}
Differentiable quantum architecture search (DQAS) is a gradient-based framework to design quantum circuits automatically in the NISQ era. It was motivated by such as low fidelity of quantum hardware, low flexibility of circuit architecture, high circuit design cost, barren plateau (BP) problem, and periodicity of weights. People used it to address error mitigation, unitary decomposition, and quantum approximation optimization problems based on fixed datasets. Quantum reinforcement learning (QRL) is a part of quantum machine learning and often has various data. QRL usually uses a manually designed circuit. However, the pre-defined circuit needs more flexibility for different tasks, and the circuit design based on various datasets could become intractable in the case of a large circuit. The problem of whether DQAS can be applied to quantum deep Q-learning with various datasets is still open. The main target of this work is to discover the capability of DQAS to solve quantum deep Q-learning problems. We apply a gradient-based framework DQAS on reinforcement learning tasks and evaluate it in two different environments - cart pole and frozen lake. It contains input- and output weights, progressive search, and other new features. The experiments conclude that DQAS can design quantum circuits automatically and efficiently. The evaluation results show significant outperformance compared to the manually designed circuit. Furthermore, the performance of the automatically created circuit depends on whether the super-circuit learned well during the training process. This work is the first to show that gradient-based quantum architecture search is applicable to QRL tasks.
\end{abstract}

\begin{IEEEkeywords}
DQAS, QRL, QAS
\end{IEEEkeywords}

\section{Introduction\label{Introduction}}
In recent years, quantum computing has developed rapidly and has achieved remarkable progress~\cite{arute2019quantum}. Diverse quantum algorithms were proposed in different fields~\cite{alam2021quantum, chen2020variational, zhang2020differentiable, ma2019variational, farhi2014quantum, beer2020training}. However, current quantum hardware has limitations on the number of qubits and the quantum gates because of the relatively low quantum gate fidelity on the deep quantum circuit~\cite{preskill2018quantum}. These noisy intermediate-scale quantum (NISQ) devices can not reach fault tolerance in the near future~\cite{preskill2018quantum}. In the NISQ era, variational quantum algorithms (VQAs) are considered the leading strategy~\cite{cerezo2021variational}. It has shown the potential to improve ML performance in quantum supervised learning (QSL)~\cite{alam2021quantum, farhi2018classification, havlivcek2019supervised, mitarai2018quantum, schuld2020circuit}, quantum unsupervised learning (QUL)~\cite{amin2018quantum, chakrabarti2019quantum, coyle2020born, zoufal2021variational} and quantum reinforcement learning (QRL)~\cite{chen2020variational, chen2022variational, jerbi2021quantum, skolik2022quantum} on NISQ devices.

Most studies in QRL focused on the framework with a predefined circuit architecture. These architectures were inspired by mathematical models and constructed through manual experience. It would take much time to design an architecture from the experts, especially when the number of qubits is increasing and the circuit is getting deeper and deeper. Besides, the lack of flexibility of predefined architecture and the hardware constraints in the NISQ era limit us from solving a QRL task efficiently. Therefore, we aim to solve these issues by automating the architecture search for QRL. This work applies quantum deep Q-learning specifically, which is an off-policy QRL algorithm. It was proposed to solve RL environments (cart pole and frozen lake) with quantum computer~\cite{jerbi2021quantum}. The circuit architecture they used is viewed as the baseline in this work.

Differentiable quantum architecture search (DQAS) is a general gradient-based framework for automating quantum circuit design problems. It has shown its effects on error mitigation and rediscovery of quantum circuits in the NISQ~\cite{zhang2020differentiable, zhang2021neural}. All DQAS previously solved problems have a specific condition with a fixed dataset, in which the losses are strongly correlated to the objective (e.g., fidelity or ground state).

However, this can not be satisfied in quantum deep Q-learning. The dataset in quantum deep Q-learning strongly depends on the current learned experiences and varies by increasing training epochs. There is no guarantee that a loss decline will lead to good rewards~\cite{miao2022differentiable}. Here is a question: Would DQAS apply to quantum deep Q-learning?

In this work, we try to apply a gradient-based framework of DQAS to reinforcement learning tasks. It is motivated by the progress toward realizing quantum advantage by utilizing DQAS in QRL. We summarise our contributions as follows:
\begin{enumerate}

\item We conduct sufficient experiments to evaluate the proposed algorithm in two RL environments based on various super-circuits. We demonstrate the success of newly discovered architectures in both environments. 

\item We also show that the discovered architecture will have a bad performance by evaluation if the corresponding super-circuit does not learn well during the training process.

\item We test the newly found architecture on a simulator of a noisy device.

\end{enumerate}

This work includes five chapters. In chapter~\ref{Introduction}, the motivation and overview of the study are introduced. Chapter~\ref {Background and Related Work} reviews two related works. Chapter~\ref{Method} shows our method. The experimental results are analyzed in chapter~\ref{Experiments}. The last chapter \ref{Conclusion and Outlook} contains conclusions and directions for future work.

\section{Related work\label{Background and Related Work}}

\subsection{DQAS}
The quantum architecture search (QAS) constructs a quantum circuit automatically. Its goal can be described as
\begin{align}
    (\theta^*,a^*)=\argmin_{\theta \in \mathcal{C},a\in S}\mathcal{L}(\theta, a, \mathcal{Z},\epsilon_a)\quad,
\end{align} where $\mathcal{L}$ represents an objective function, $\theta$ denotes the parameters of an ansatz $a$, $\mathcal{C}$ is a constraint set, $\mathcal{Z}$ represents the input, and $\epsilon_a$ denotes the quantum system noise.

        As one of the QAS algorithms, DQAS is a hybrid quantum framework~\cite{zhang2020differentiable}. It has one super-circuit including an operation pool $\mathcal{O}$, $p$ placeholders, architecture parameters $\alpha_p$, and circuit weights $\theta$ with dimension $p\times|\mathcal{O}|\times l$, where $l$ is the largest number of gate parameters. It is inspired by differentiable architecture search (DARTS) and designs a circuit by placing multiple operation candidates in a placeholder. Given some placeholders and an operations pool, people usually test all combinations of placeholders and operation candidates individually to get the best-performing architecture. The search space is thus discrete. However, DQAS gives weight to each operation candidate within a placeholder, and people refer to the normalized weights as operation probabilities. The probability of a circuit with a specific architecture is then defined as the product of probabilities of its component's operations. DQAS makes the search space continuous through this probability model. It samples a batch of circuit architecture candidates to calculate the global loss, which is a sum of the loss of each batch member and represented as
        \begin{align}
            \mathcal{L}=\sum_{k\sim P(k,\alpha)}\frac{P(k,\alpha)}{\sum_{k^{'}\sim P(k,\alpha)}P(k^{'},\alpha)}L_k(\theta)\quad,
        \end{align}
        where
        \begin{align}
            P(k,\alpha)=\prod_i^p \frac{e^{\alpha_{ij}}}{\sum_k e^{\alpha_{ik}}}\quad,
        \end{align} $k$ denotes the sampled architecture from the probability model $P(k,\alpha)$.
        The architecture and circuit parameters are optimized using gradient descent~\cite{zhang2020differentiable} with a global loss function.
        
        In practice, DQAS can compose a quantum circuit for error mitigation and optimization problems~\cite{zhang2020differentiable}.

\subsection{QRL}
Similarly to the classical deep Q-learning algorithm, quantum deep Q-learning uses a variational quantum circuit VQC as a counterpart of DQN. VQC here is called Quantum-DQN (QDQN). One observation state is input for the encoding block, and measurement provides output. Trainable parameters of QDQN correspond to the parameters of DQN. There is a replay memory to manage observations. The Q-values are calculated by
\begin{align}
Q(s,a)=\frac{1}{2}(\langle0^{\otimes n}|U_{\theta}^{\dagger}(s)O_a U_{\theta}(s)|0^{\otimes n}\rangle+1)
\end{align} in range of $[0, 1]$ for each action $a$ by corresponding observables $O_a$. The local loss function is represented as
        \begin{align}
            L_i(\theta_i)&=\mathbb{E}_{(s,a,r,s')\sim ER}[(r\nonumber\\
            &+\gamma \text{max}_{a'}Q^{\text{target-QDQN}}(s',a';\theta_i^-)\\
            &-Q^{\text{QDQN}}(s,a;\theta_i))^2]\quad,\nonumber
        \end{align}where $\gamma$ denotes the discount factor. The vector $(s,a,r,s')$ is one of the sampled observations including state $s$, action $a$, reward $r$ and next state $s'$. $\theta_i^-$ and $\theta_i$ denote parameters of target-QDQN and QDQN, respectively. For every $n$ epochs, the target-QDQN copies $\theta_i$ from QDQN into itself.
        
        With the help of quantum deep Q-learning, the cart-pole environment and the frozen lake environment were solved in~\cite{skolik2022quantum}. They also used data re-uploading technique input and output weights to improve training performance. We will refer to these methods in our following work as new features for DQAS.

\section{Method\label{Method}}
Algorithm~\ref{alg:algo3 RL-DQAS} shows an overview for RL-DQAS. Firstly, a super-circuit is defined, including a circuit with placeholders, an operation pool $\mathcal{O}$, and trainable shared parameters for circuit gates $\theta$ and architecture $\alpha$. Secondly, we start training with multiple agents. According to the current architecture distribution model $P$, a batch of architecture candidates is sampled from the super-circuit. Each candidate shares circuit parameters $\theta$ and calculates individual loss value $\mathcal{L}$ between the predicted and target Q-values. With a global loss, the shared architecture parameters $\alpha$ and circuit parameters $\theta$ can then be updated by gradient descent with optimizer Adam. Meanwhile, the corresponding probability is updated as well. For every $n$ iterations, the progressive search checks for each placeholder if one operation candidate and its architecture parameter should be removed by comparing the corresponding architecture distribution. The target circuit will periodically copy the circuit values of the shared parameters. 
\begin{algorithm}[ht!bp]
\caption{DQAS for RL}\label{alg:algo3 RL-DQAS}
    \begin{algorithmic}
        \State $\textbf{Step 1: Super-circuit definition: }$
        \Indent
        \State Initialize op $\mathcal{O}$ and super-circuit with $\alpha$ and $\theta$
        \State Initialize two QDQNs and environment $e$
        \EndIndent
        \State $\textbf{Step 2: Super-circuit training: }$
        \Indent
        \While{Architecture search}
        \State Sample minibatch of architectures
        \State Calculate global loss $\mathcal{L}$ via Eq:~\ref{eq:global loss}
        \State Update $\alpha_t$ and  $\theta_t$ via gradients $\nabla\mathcal{L}_\alpha$ 
        and $\nabla\mathcal{L}_\theta$ 
        \State The tar-circuit copy $\theta$ from pred-circuit periodically
        \EndWhile
        \State Circuit parameter tuning
        \State Early stop training if $e \ni r^{avg} \geq r^{max} \in e$
        \EndIndent
        \State $\textbf{Step 3: Rank training performance among agents: }$
        \Indent
        \State Take top K architectures for evaluation
        \EndIndent
        \State $\textbf{Step 4: Evaluation of architectures: }$
        \Indent
        \State Take the best performing architecture or retrain
        \EndIndent
  \end{algorithmic}
\end{algorithm}
If the architecture is fixed, the tuning phase only updates the trained circuit parameters iteratively before the reward converges to a given value. After training, we rank training performances among agents and take the architectures of the first $K$ top-performing candidates. Finally, we evaluate these architectures to find out the best-performing architecture.

\subsection{Super-circuit}
A super-circuit builds a search space using a circuit with placeholders, architecture parameters $\alpha$, and operations from operation pool $\mathcal{O}$.
The circuit consists of three blocks. The encoding block encodes the weighted input state into a quantum state. The parameterized block contains all placeholders and should be stacked to create a deep circuit. The measurement block is used to provide output value for a classical computer.

The super-circuit adopts the architecture distribution model $P(U,\alpha)$, where $U$ denotes the sequence of placeholders or an architecture candidate. This work defines operation candidates and placeholders differently than the original DQAS. Each placeholder $u_i$ covers all qubits instead of one qubit and accepts one element from the operation pool containing the working range. This way, the number of parameters for each placeholder depends on the operation type and working range, and the search space can thus be reduced by controlling the working range of operation candidates. One parameterized block can then be described as
\begin{align}
    U = \prod_{i=0}^{p} u_i(\theta_i)\quad,
\end{align}
where $u_i$ stands for unitary placeholder and can be replaced by any $o_i\in \mathcal{O}$. $\theta_i$ can vanish if $o_i$ has no trainable parameter. Furthermore, the depth of the block increases with every filled placeholder.

\subsection{Operation pool}\label{subsection: operation pool}
Operation pool $\mathcal{O}$ in size of $s=|\mathcal{O}|$ is a set of quantum gate candidates. Each operation in $\mathcal{O}$ contains the type of operation and its working range. If there is a circuit with four qubits, an operation pool can be defined with elements:
\begin{align}
 \mathcal{O}&=\{\underbrace{o_1}_{\textbf{Type}}:\underbrace{[1,2,3,4]}_{\textbf{Working range}}, o_2:[1,2,3,4],\\
 &o_3:[1,2,3], o_4:[2,3,4],E:[1,2,3,4]\}\quad.\nonumber
\end{align}
The $o_i$ denotes the type of operation, and it could be any 1-qubit gate or multi-qubits gate (e.g., \texttt{RZ}, \texttt{U3}, or \texttt{CNOT}). The corresponding array shows the range in which the operation works. For example, if operation $o_1=$ \texttt{CNOT} is selected for one placeholder, four \texttt{CNOT} gates will work on all qubits with ring connections. This work defines two operation pools $op3$ and $op4$.  $op3$ contains candidates such as \texttt{ry}, \texttt{rz}, \texttt{cz}, \texttt{cnot} and identity on all qubits, and there are candidates \texttt{ry} and \texttt{rz} both further working on $\{[1,2,3],[2,3,4],[1,2],[2,3],[3,4]\}$. $op4$ contains almost the same candidates but has no \texttt{cz}, neither \texttt{ry} nor \texttt{rz} on $\{[1,2],[2,3],[3,4]\}$.

\subsection{Objectives and gradients}
In this work, the objective function takes two Q-values as inputs, created from two quantum circuits (predicting circuit and target circuit). The predicted Q-value should gradually converge to the target Q-value. The local objective $L(U,\theta)$ is calculated as the mean squared error (MSE) between predicted Q-values and target Q-values
\begin{align}
    L(U,\theta) = (Q^{\text{predicted}}_{U, \theta}-Q^{\text{target}}_{U, \theta^{\ast}})^2\quad,\label{eq:local loss}
\end{align}
where $U$ stands for circuit architecture and $\theta$ represents circuit parameter tensor. The global objective function is defined as the total sum of local objectives according to the architecture distribution model $P(U, \alpha)$:
\begin{align}
    \mathcal{L}=\sum_{U\sim P(U,\alpha)}L(U,\theta)\label{eq:global loss}\quad.
\end{align}
The Q-value for action $a\in A$ is calculated by the circuit measurement with the corresponding observable $O_a$. In order to compare Q-values easily, we shift and scale the measurement result by $1$ and $\frac{1}{2}$ respectively.

We will iteratively update the parameters by gradient descent, including the circuit parameter tensor $\theta$, the architecture parameter matrix $\alpha$, and the input weights $w^{in}$, the output weights $w^{out}$ if needed. Since the circuit tensor $\theta$ is independent of the architecture distribution, its gradient takes the form
\begin{align}
    \nabla_{\theta} \mathcal{L}=\sum_{U\sim P(U,\alpha)}\nabla_{\theta} L(U,\theta)\quad.\label{eq:gradient-theta}
\end{align}

\begin{figure}[t!]
    \centering
    \subfloat[Frozen Lake\label{fig:fl 1 best structure}]{%
        \includegraphics[width=0.4\linewidth]{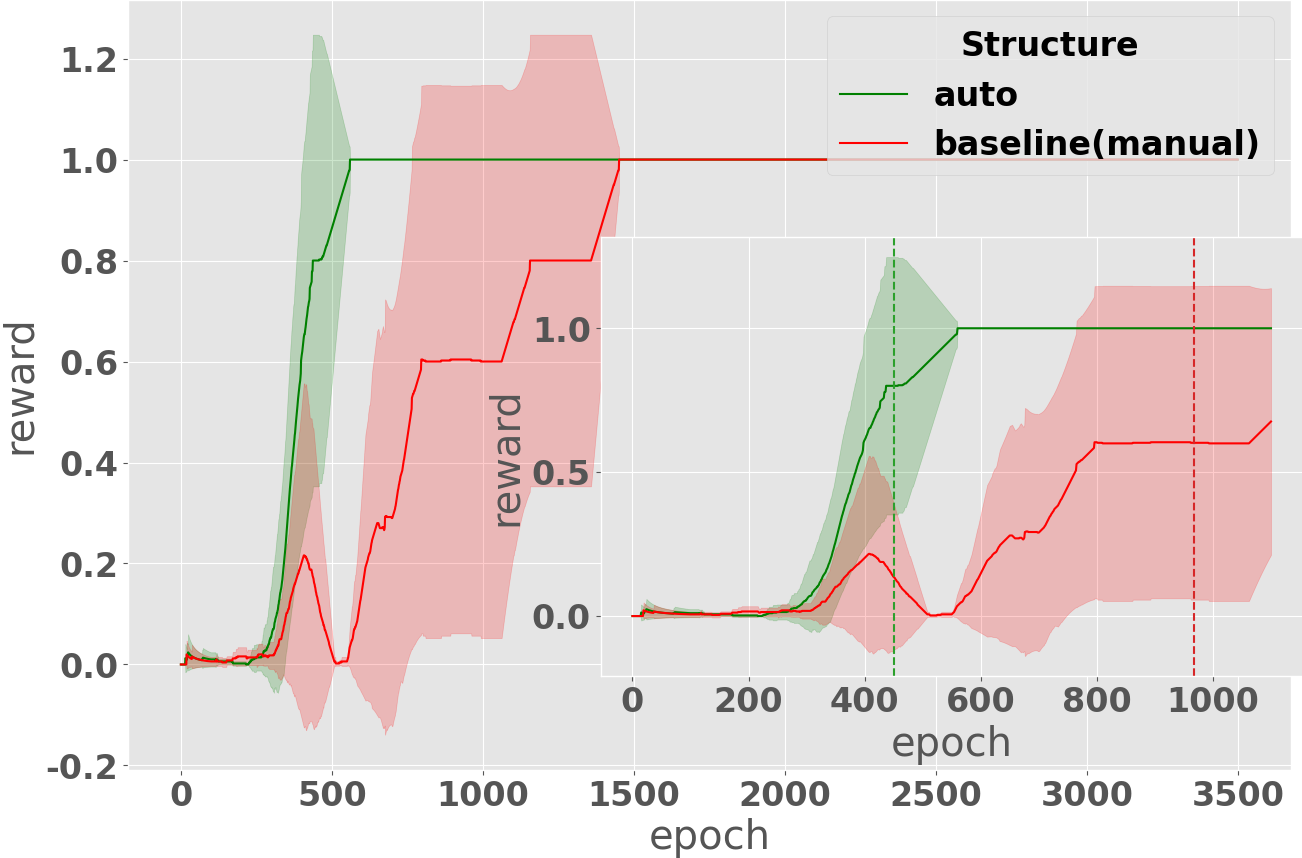}}
    \quad
    \subfloat[Cart Pole\label{fig:cp 1 best structure}]{%
       \includegraphics[width=0.4\linewidth]{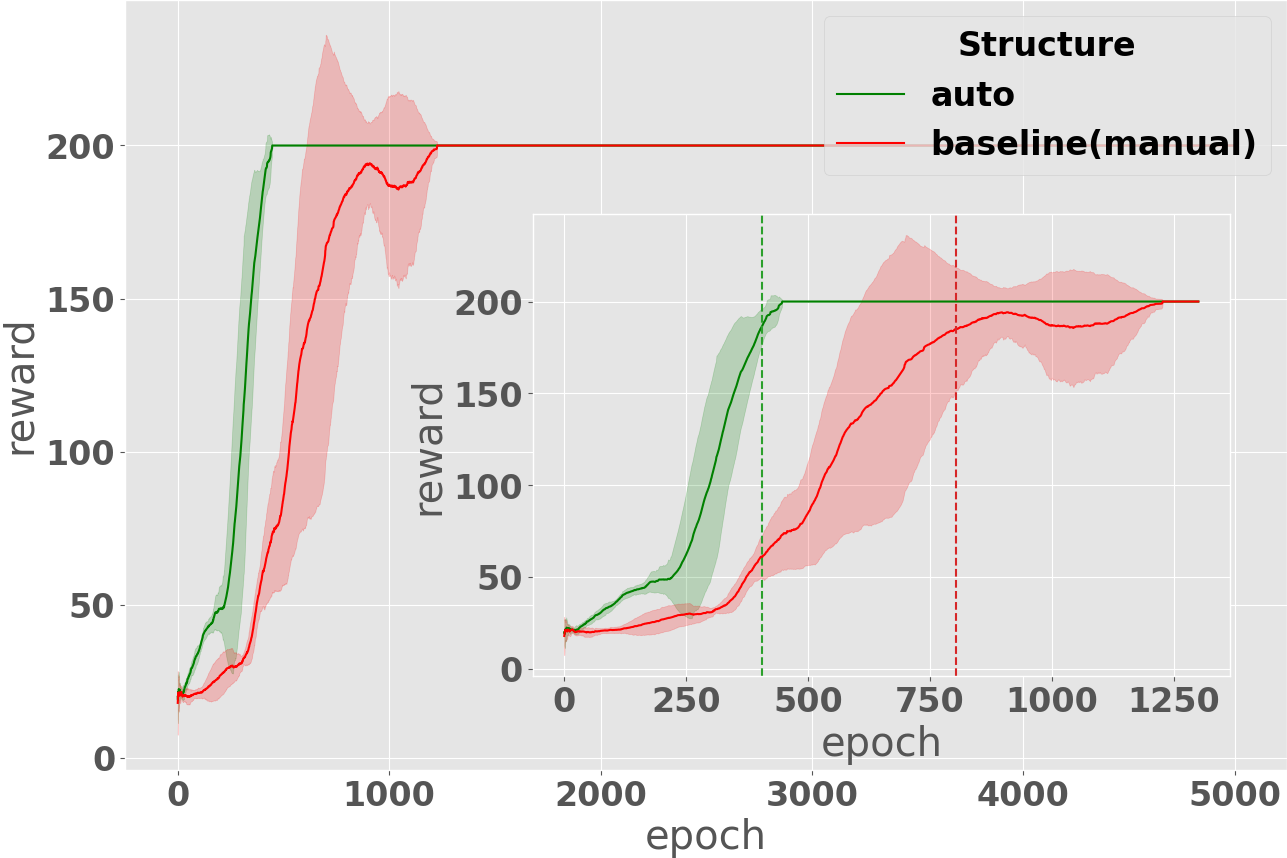}}
    \\
    \subfloat[Architecture Auto-fl-op3]{
    \includegraphics[width=0.4\linewidth]{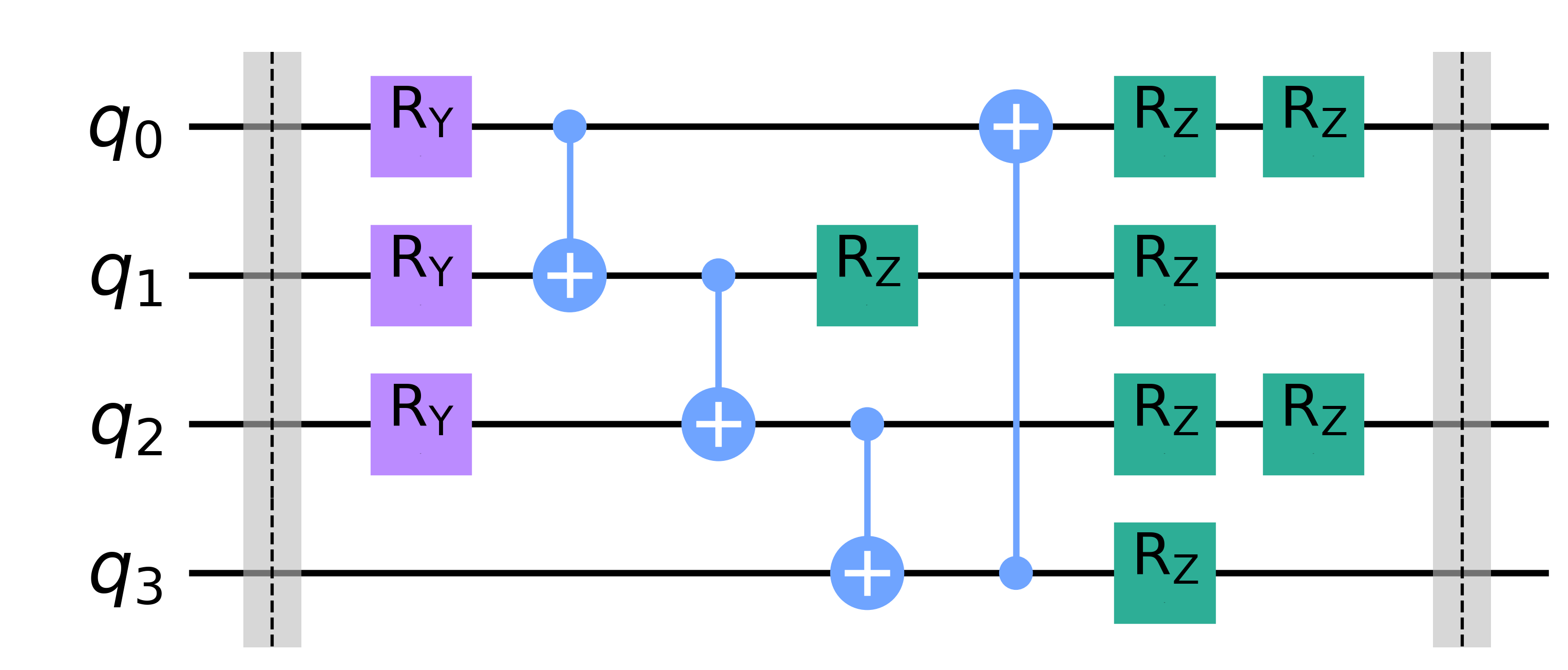}
    \label{fig:Architecture Auto-op3 cp) fl}
    }
    \quad
    \subfloat[Architecture Auto-cp-op4]{
    \includegraphics[width=0.4\linewidth]{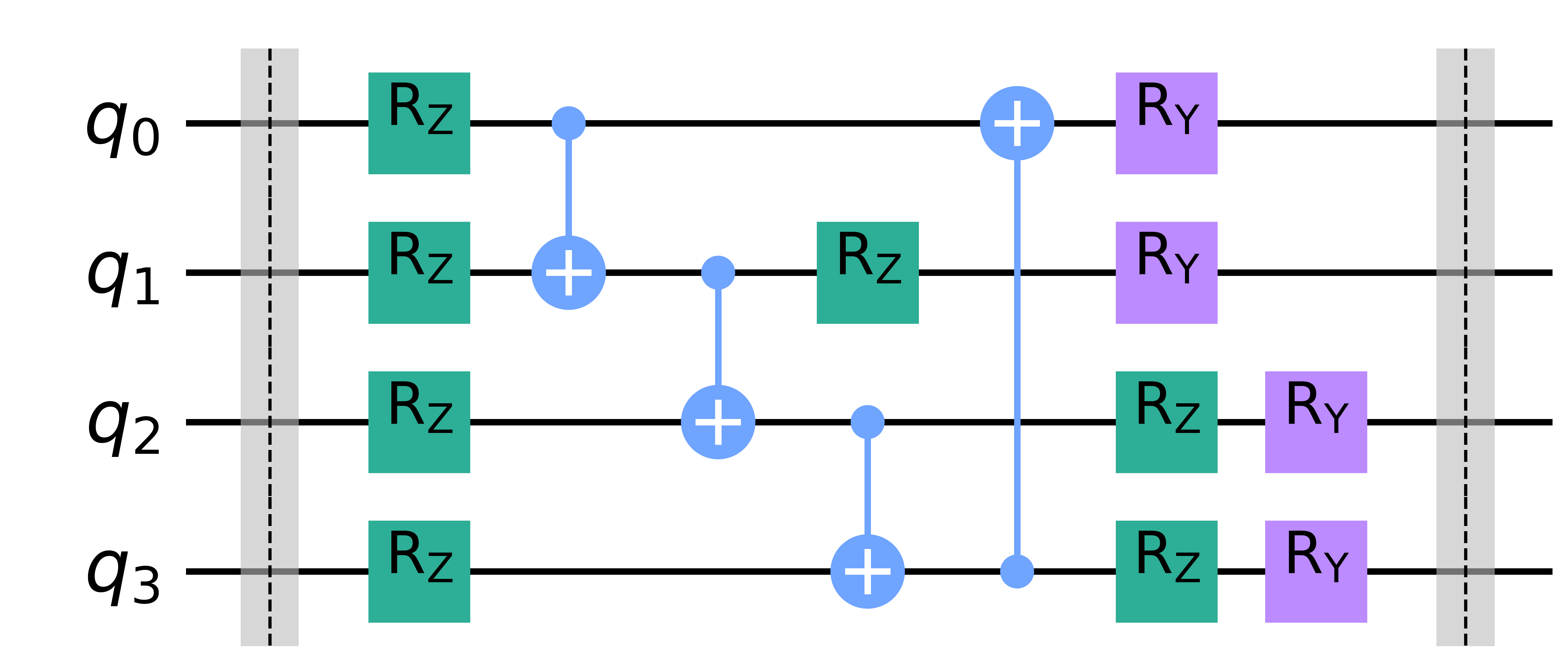}
    \label{fig:Architecture Auto-op4 cp)}
    }
  \caption{Evaluation and information of newly found architectures on the simulator without noise. The results are averaged over five agents, where each
agent has five parameterized blocks.}
  \label{fig:best structures} 
\end{figure}
In practice, this gradient can be calculated by the parameter shift rules~\cite{crooks2019gradients, mitarai2018quantum, schuld2019evaluating}. The gradient of the architecture parameter matrix $\alpha$ is related to the architecture distribution and calculated as described in DQAS~\cite{zhang2020differentiable}.

\section{Experiments and results\label{Experiments}}

\subsection{Experiment setting}
In this chapter, we conduct two benchmark RL experiments (e.g., Cart Pole v0~\cite{openai-gym-cp} and Frozen Lake v0\cite{openai-gym-fl}) from OpenAI Gym.
This work uses four qubits and repeats one parameterized block five times to create the super-circuit. Each block has four placeholders. We select the same encoding scheme and observables used in quantum deep Q-learning~\cite{skolik2022quantum} for both experiments, and their manually designed circuit architecture is referred to as our baseline. The baseline circuit has five parameterized blocks, each composed of three operations columns. Each column covers all qubits, followed by \texttt{ry}, \texttt{rz}, \texttt{cz}.

\subsection{Results and discussion}
We first focus on the most surprising result shown in Fig.~\ref{fig:best structures}. The automatically-designed architecture (green lines) improves the performance by about $200\%$ compared to the baseline, decreasing the average solving point from around $800$ to $400$ for the cart pole and $1000$ to $400$ for the frozen lake. Furthermore, the shaded standard deviation of this architecture is much smaller than the baseline, showing the homogeneous training performance of agents.

The best architecture Auto-op4 shown in Fig.~\ref{fig:Architecture Auto-op4 cp)} derives from the super-circuit \texttt{op4} during the training process. By manual design, \texttt{CNOT} gates are placed at a block's front or back end, and people rarely reuse gates. In contrast, the machine places the \texttt{CNOT} gates in the middle of the parameterized block and then chooses to add some additional \texttt{RZ} gates on part of qubits. This design could provide more control for the z-axis on these selected qubits and improve the mapping between agent state and agent action. Although the reason for this selection is unknown, in practice, this design improves the training performance significantly and is hard to mimic.

As shown in Fig~\ref{fig:Architecture Auto-op3 cp) fl}, the architecture has redundant gates - \texttt{RZ} gates at the rear, which is distinct from manually designed architectures. These redundant gates might improve the training performance through stacking parameterized blocks.

\begin{figure}[t!]
    \centering
    \includegraphics[width=0.5\linewidth]{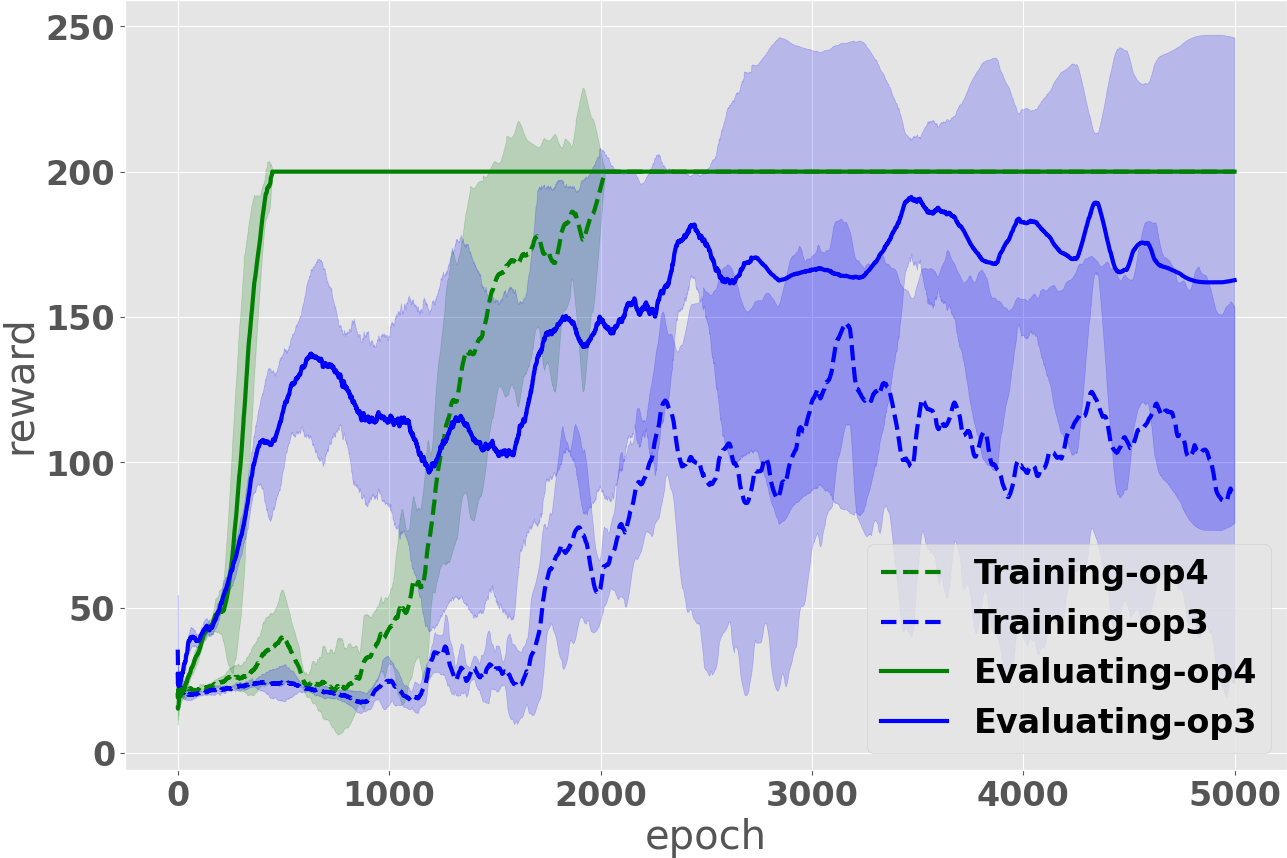}
    \caption{Relationship between training and evaluation.}
    \label{fig:Training vs. Evaluating cp}
\end{figure}

We now show the performance of discovered architectures and corresponding super-circuits in Fig.~\ref{fig:Training vs. Evaluating cp}. While the super-circuit op4 (dashed green line) learns well during the training process, the super-circuit op3 (dashed blue line) has a bad performance, which might be caused by instability of RL or bad super-circuit definition. The difference in learning performance between these two super-circuits results in the evaluation difference between corresponding designed architectures. Additionally, the circuit architectures of the five agents do not show convergency, while the training performance converges to the top. Different architectures could perform similarly. The performance and architectural similarity could be discussed in the future. Consequently, the performance of the designed circuit depends on whether the corresponding super-circuit learns well during the training process. A badly performing architecture can be ascribed to the failure of the training process of the super-circuit.

    

\begin{figure}[t!]
    \centering
  \subfloat[op3\label{fig:real simulator aer ibm_quito fl op3}]{%
       \includegraphics[width=0.4\linewidth]{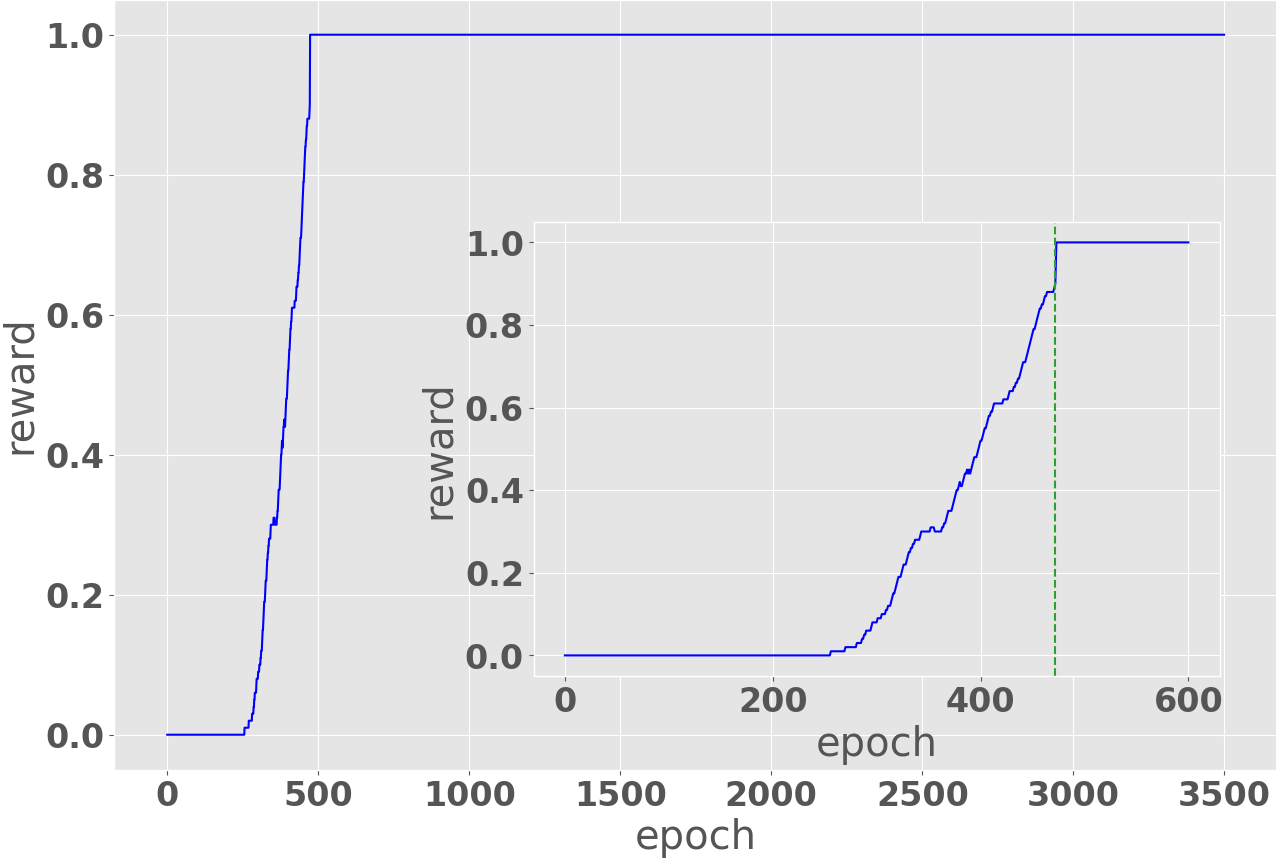}}
    \quad
  \subfloat[op4\label{fig:real simulator aer ibm_quito fl op4}]{%
        \includegraphics[width=0.4\linewidth]{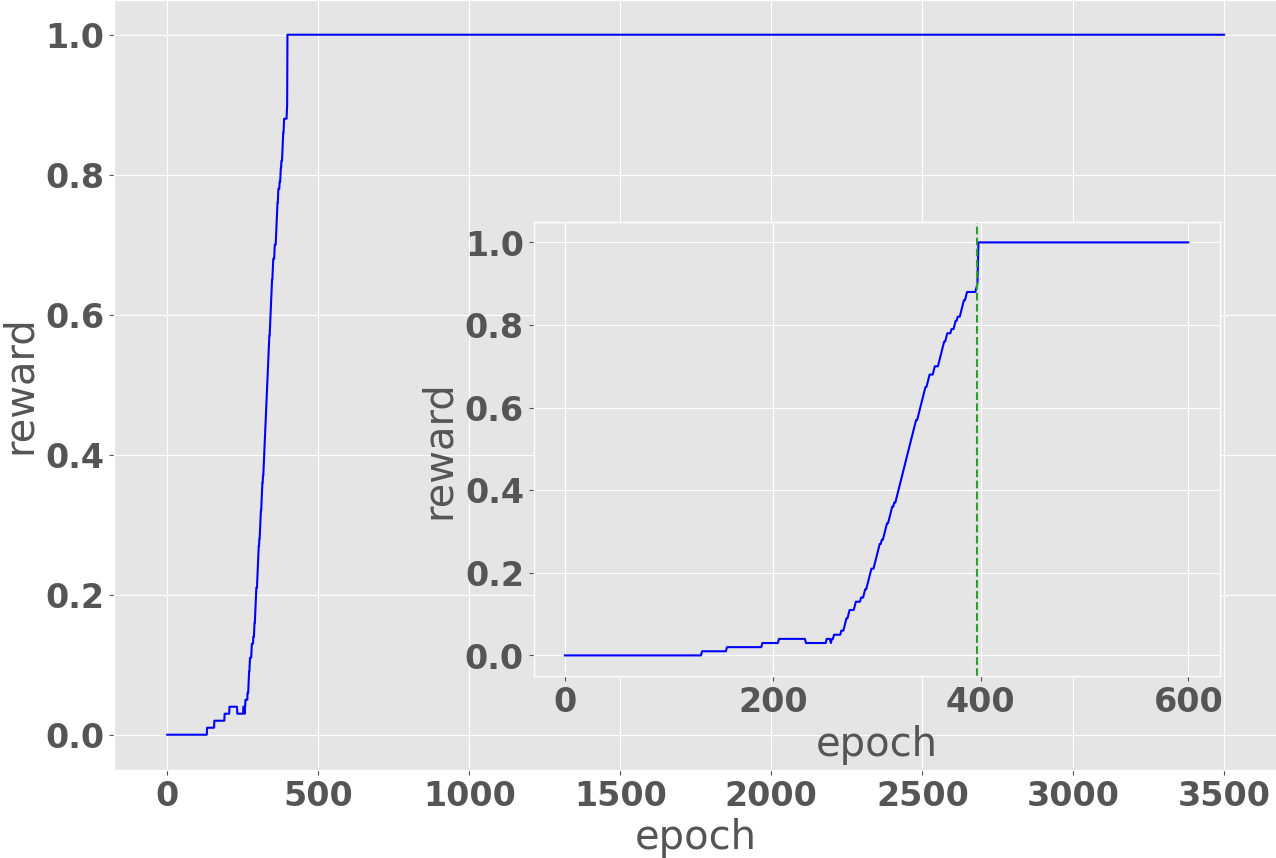}}
    \\
  \caption{Evaluation in the frozen lake environment. (IBM noisy simulator qiskit-aer "ibmq-quito").}
  \label{fig:real device simulator aer ibmq_quito} 
\end{figure}

In addition, as shown in Fig.~\ref{fig:real device simulator aer ibmq_quito}, we investigate whether these two architectures work in the same environment on the noisy device. Due to unpredictable delays of real hardware, we utilize the noisy simulator to evaluate two architectures in the frozen lake environment. We are surprised that the newly discovered circuit architectures performed well on the noise model, although one is designed for another experiment. It will also encourage further investigation into a general architecture based on the gradient-based QAS for multiple QRL tasks. In the next chapter, we will conclude.

\section{Conclusion and Outlook\label{Conclusion and Outlook}}
In this work, we apply DQAS for QRL. The results show that it can design quantum circuit architectures in different RL environments (cart pole and frozen lake) efficiently and automatically. We successfully searched with different super-circuits, and the newly discovered architectures differed from and outperformed the manually designed architecture. Furthermore, we show that the evaluation depends on the training performance. We find a general architecture in the noisy model for the frozen lake. This work is the first to show that gradient-based QAS applies to QRL tasks.

This work uses a shallow super-circuit and mini-learning-step for the training process. However, better architecture might hide in a deep super-circuit. Furthermore, we could find some common circuit architecture features for multiple RL tasks.
\section*{Acknowledgment}
The project of this workshop paper is based on was supported with funds from the German Federal Ministry of Education and Research in the funding program Quantum Reinforcement Learning for industrial Applications (QLindA) - under project number 13N15644. The sole responsibility for the paper’s contents lies with the authors.

\bibliographystyle{ieeetr} 
\bibliography{mybib}

\begin{thebibliography}{10}

\bibitem{arute2019quantum}
F.~Arute, K.~Arya, R.~Babbush, D.~Bacon, J.~C. Bardin, R.~Barends, R.~Biswas,
  S.~Boixo, F.~G. Brandao, D.~A. Buell, {\em et~al.}, ``Quantum supremacy using
  a programmable superconducting processor,'' {\em Nature}, vol.~574, no.~7779,
  pp.~505--510, 2019.

\bibitem{alam2021quantum}
M.~Alam, S.~Kundu, R.~O. Topaloglu, and S.~Ghosh, ``Quantum-classical hybrid
  machine learning for image classification (iccad special session paper),''
  {\em arXiv preprint arXiv:2109.02862}, 2021.

\bibitem{chen2020variational}
S.~Y.-C. Chen, C.-H.~H. Yang, J.~Qi, P.-Y. Chen, X.~Ma, and H.-S. Goan,
  ``Variational quantum circuits for deep reinforcement learning,'' {\em IEEE
  Access}, vol.~8, pp.~141007--141024, 2020.

\bibitem{zhang2020differentiable}
S.-X. Zhang, C.-Y. Hsieh, S.~Zhang, and H.~Yao, ``Differentiable quantum
  architecture search,'' {\em arXiv preprint arXiv:2010.08561}, 2020.

\bibitem{ma2019variational}
Y.~Ma, V.~Tresp, L.~Zhao, and Y.~Wang, ``Variational quantum circuit model for
  knowledge graph embedding,'' {\em Advanced Quantum Technologies}, vol.~2,
  no.~7-8, p.~1800078, 2019.

\bibitem{farhi2014quantum}
E.~Farhi, J.~Goldstone, and S.~Gutmann, ``A quantum approximate optimization
  algorithm,'' {\em arXiv preprint arXiv:1411.4028}, 2014.

\bibitem{beer2020training}
K.~Beer, D.~Bondarenko, T.~Farrelly, T.~J. Osborne, R.~Salzmann,
  D.~Scheiermann, and R.~Wolf, ``Training deep quantum neural networks,'' {\em
  Nature communications}, vol.~11, no.~1, pp.~1--6, 2020.

\bibitem{preskill2018quantum}
J.~Preskill, ``Quantum computing in the nisq era and beyond,'' {\em Quantum},
  vol.~2, p.~79, 2018.

\bibitem{cerezo2021variational}
M.~Cerezo, A.~Arrasmith, R.~Babbush, S.~C. Benjamin, S.~Endo, K.~Fujii, J.~R.
  McClean, K.~Mitarai, X.~Yuan, L.~Cincio, {\em et~al.}, ``Variational quantum
  algorithms,'' {\em Nature Reviews Physics}, vol.~3, no.~9, pp.~625--644,
  2021.

\bibitem{farhi2018classification}
E.~Farhi and H.~Neven, ``Classification with quantum neural networks on near
  term processors,'' {\em arXiv preprint arXiv:1802.06002}, 2018.

\bibitem{havlivcek2019supervised}
V.~Havl{\'\i}{\v{c}}ek, A.~D. C{\'o}rcoles, K.~Temme, A.~W. Harrow, A.~Kandala,
  J.~M. Chow, and J.~M. Gambetta, ``Supervised learning with quantum-enhanced
  feature spaces,'' {\em Nature}, vol.~567, no.~7747, pp.~209--212, 2019.

\bibitem{mitarai2018quantum}
K.~Mitarai, M.~Negoro, M.~Kitagawa, and K.~Fujii, ``Quantum circuit learning,''
  {\em Physical Review A}, vol.~98, no.~3, p.~032309, 2018.

\bibitem{schuld2020circuit}
M.~Schuld, A.~Bocharov, K.~M. Svore, and N.~Wiebe, ``Circuit-centric quantum
  classifiers,'' {\em Physical Review A}, vol.~101, no.~3, p.~032308, 2020.

\bibitem{amin2018quantum}
M.~H. Amin, E.~Andriyash, J.~Rolfe, B.~Kulchytskyy, and R.~Melko, ``Quantum
  boltzmann machine,'' {\em Physical Review X}, vol.~8, no.~2, p.~021050, 2018.

\bibitem{chakrabarti2019quantum}
S.~Chakrabarti, Y.~Huang, T.~Li, S.~Feizi, and X.~Wu, ``Quantum wasserstein
  generative adversarial networks,'' {\em arXiv preprint arXiv:1911.00111},
  2019.

\bibitem{coyle2020born}
B.~Coyle, D.~Mills, V.~Danos, and E.~Kashefi, ``The born supremacy: quantum
  advantage and training of an ising born machine,'' {\em npj Quantum
  Information}, vol.~6, no.~1, pp.~1--11, 2020.

\bibitem{zoufal2021variational}
C.~Zoufal, A.~Lucchi, and S.~Woerner, ``Variational quantum boltzmann
  machines,'' {\em Quantum Machine Intelligence}, vol.~3, no.~1, pp.~1--15,
  2021.

\bibitem{chen2022variational}
S.~Y.-C. Chen, C.-M. Huang, C.-W. Hsing, H.-S. Goan, and Y.-J. Kao,
  ``Variational quantum reinforcement learning via evolutionary optimization,''
  {\em Machine Learning: Science and Technology}, vol.~3, no.~1, p.~015025,
  2022.

\bibitem{jerbi2021quantum}
S.~Jerbi, L.~M. Trenkwalder, H.~P. Nautrup, H.~J. Briegel, and V.~Dunjko,
  ``Quantum enhancements for deep reinforcement learning in large spaces,''
  {\em PRX Quantum}, vol.~2, no.~1, p.~010328, 2021.

\bibitem{skolik2022quantum}
A.~Skolik, S.~Jerbi, and V.~Dunjko, ``Quantum agents in the gym: a variational
  quantum algorithm for deep q-learning,'' {\em Quantum}, vol.~6, p.~720, 2022.

\bibitem{zhang2021neural}
S.-X. Zhang, C.-Y. Hsieh, S.~Zhang, and H.~Yao, ``Neural predictor based
  quantum architecture search,'' {\em Machine Learning: Science and
  Technology}, vol.~2, no.~4, p.~045027, 2021.

\bibitem{miao2022differentiable}
Y.~Miao, X.~Song, J.~D. Co-Reyes, D.~Peng, S.~Yue, E.~Brevdo, and A.~Faust,
  ``Differentiable architecture search for reinforcement learning,'' in {\em
  First Conference on Automated Machine Learning (Main Track)}, 2022.

\bibitem{crooks2019gradients}
G.~E. Crooks, ``Gradients of parameterized quantum gates using the
  parameter-shift rule and gate decomposition,'' {\em arXiv preprint
  arXiv:1905.13311}, 2019.

\bibitem{schuld2019evaluating}
M.~Schuld, V.~Bergholm, C.~Gogolin, J.~Izaac, and N.~Killoran, ``Evaluating
  analytic gradients on quantum hardware,'' {\em Physical Review A}, vol.~99,
  no.~3, p.~032331, 2019.

\bibitem{openai-gym-cp}
G.~Brockman, V.~Cheung, L.~Pettersson, J.~Schneider, J.~Schulman, J.~Tang, and
  W.~Zaremba, ``Openai gym, wiki, cartpole v0.''
  \url{https://github.com/openai/gym/wiki/CartPole-v0}, 2016.

\bibitem{openai-gym-fl}
G.~Brockman, V.~Cheung, L.~Pettersson, J.~Schneider, J.~Schulman, J.~Tang, and
  W.~Zaremba, ``Openai gym, wiki, frozenlake v0.''
  \url{https://github.com/openai/gym/wiki/FrozenLake-v0}, 2016.

\end{thebibliography}

\end{document}